\documentclass[%
 reprint,
 twocolumn,
 amsmath,amssymb,
 aps,
]{revtex4}

\pdfoutput=1
\usepackage[utf8]{inputenc}
\usepackage[english]{babel}
\usepackage[T1]{fontenc}
\usepackage{amsmath}
\usepackage{amsfonts}
\usepackage{amssymb}
\usepackage{hyperref}
\usepackage{tikz}
\usepackage{lipsum}
\usepackage{graphicx}
\usepackage{color}
\usepackage{verbatim}
\usepackage{lipsum}
\usepackage{dsfont}
\usepackage{upgreek}
\usepackage{ulem} 

\begin{document}


\title{A Deterministic Single Ion Fountain}

\author{Felix Stopp}
\email{felstopp@uni-mainz.de}
 \homepage{http://www.quantenbit.physik.uni-mainz.de/}
 \affiliation{QUANTUM, Institut f\"ur Physik, Universit\"at Mainz, Staudingerweg 7, 55128 Mainz, Germany}
\author{Henri Lehec}
\affiliation{QUANTUM, Institut f\"ur Physik, Universit\"at Mainz, Staudingerweg 7, 55128 Mainz, Germany}
\author{Ferdinand Schmidt-Kaler}
\affiliation{QUANTUM, Institut f\"ur Physik, Universit\"at Mainz, Staudingerweg 7, 55128 Mainz, Germany}%

\date{\today}

\begin{abstract}
We present an alternative approach for interconnecting trapped ion processor nodes by a deterministic single ion transfer out of the trap, into a free space trajectory, followed by recapture in the trapping potential. Our experimental realization yields a success probability of 95.1$\%$, namely 715 out of 752 extracted ions are retrapped, cooled and observed after a transport distance of $110\,\mathrm{mm}$ and a time of flight of $7\,\upmu\mathrm{s}$. Based on the near-unity operation success, we discuss its application for scalable ion trap quantum computing and advanced quantum sensing. 
\end{abstract}

\maketitle

\section{Introduction}
Trapped ions are among the leading platforms for quantum computing and quantum simulations. The basis of this success is the full control over the internal electronic states of trapped ions to encode qubits \cite{haffner2008quantum,blatt2012quantum}. The external degrees of freedom, forming collective eigenmodes due to the long range Coulomb interaction in ion crystals, are employed to drive quantum gate operations and to generate multi-particle entanglement \cite{haffner2005scalable, monroe2021programmable}. Spatial control of trapped ions and ion crystals in a segmented trap \cite{bowler2012coherent,ruster2014experimental,kaufmann2017fast,lekitsch2017blueprint} provides the basis for a prominent shuttling based approach to multi-qubit scalability \cite{kielpinski2002architecture,hilder2021fault, dumitrescu2021realizing}. Alternatively, quantum information may be converted into photons for building networks of quantum processors \cite{monroe2013scaling, monroe2014large,bock2018high,takahashi2020strong}. The interconnectivity of ion-based quantum processors nevertheless remains a significant challenge.

The deterministic transport of a single ion carrying a qubit  between two different ion traps constitutes so far unexplored pathway for conveying quantum information \cite{lekitsch2017blueprint}. Unlike the photon-mediated approach one can reach unity-efficiency, while realizing the transport at time scales much faster than common gate times, or times required for ion qubit register reordering \cite{kaufmann2017scalable, wan2020ion, kaushal2020shuttling}. So far, spatial control of the ion has been limited to the trapping region. In this work we are extending this control into the free space region, realizing an ion fountain. We will demonstrate the required methods and their optimization for the extraction of an ion from a trapping potential to a free space region, the steering of its trajectory and its recapture in the trap. We see a high application potential of our method for the development of networking between ion quantum processors. The control over a single ion in free space can, however, also enable new quantum sensing applications \cite{ruster2017entanglement}. 

\begin{figure}[h!]
\includegraphics[width=0.49\textwidth]{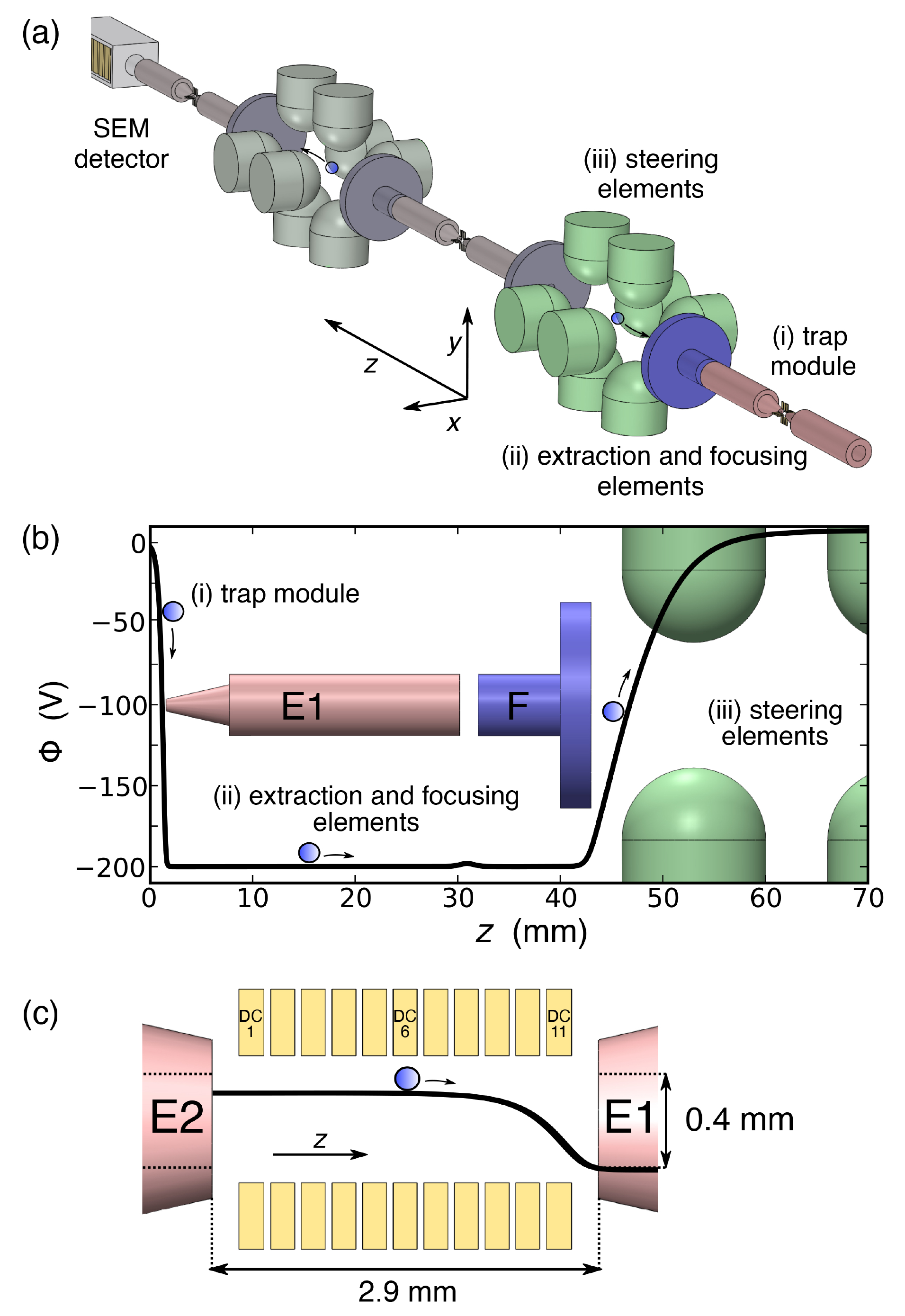}
\caption{(a) Sketch of the overall apparatus made up from a stack of (i) trap modules (yellow), (ii) extraction $\&$ focusing module (red/blue) and (iii) steering module (green). Ions are moved back and forth along the z-axis. The extension for trap-to-trap connectivity and detection of single ions in a secondary electron multiplier (SEM) is not used in the actual experiments (grey). (b) Detailed view of arrangement of modules (i) to (iii) and the axial potential (black line) along z-axis. (c) Trapping module with segmented X-blades and pierced endcaps E1 and E2 for extraction and recapture. The parameters used for the calculation of the axial potential are $U^{(6)}_\mathrm{seg}=-0.6\,\mathrm{V}$, otherwise $0\,\mathrm{V}$. Furthermore we apply $U_\mathrm{E}^{(1)}=U_\mathrm{F}=-200\,\mathrm{V}$, $U_\mathrm{R}=+7.5\,\mathrm{V}$, $U_\mathrm{E}^{(2)}$=0.}
\label{fig:1}
\end{figure}

In this letter, we first describe the setup and experimental sequence. Then we show the optimization of  parameters, starting with numerical simulations, followed by a series of calibration experiments. We employ a specialized linear ion trap setup for the experiments: ions are extracted and - instead of using a second identical linear trap downstream - we reflect the ion trajectories and capture the ions entering from free space in the linear trap where they started initially.  In doing so, we realize an electrostatic cat's eye ion reflector. We believe this ion fountain experiment fully establishes all techniques, which are required for trap-to-trap ion transport. The procedure of ion extraction, free flight, and single ion recapture, achieves a success probability of more than $95\,\%$. As an outlook, we sketch future improvements and outline various applications of the established single ion free-space transports.  

\section{Experimental setup}
The overall setup used in this work is made up of three basic modules, (i) linear segmented ion trap module, (ii) extraction and focusing elements, and (iii) electrostatic steering elements, shown in  \hyperref[fig:1]{Fig. \ref{fig:1}(a)}, which can be stacked freely in any combination onto each other as they are held mechanically stable along the $z$-axis in tightly fitting titanium mounts.  Not used in the setup is an Einzel-lens module which may serve for tight ion focusing \cite{jacob2016transmission, groot2019deterministic}.  

The trap module (i) contains a segmented Paul trap in X-shape configuration~\cite{jacob2016transmission}. Blades are fabricated from laser-cut gold-covered alumina wafers and each of the 11 DC segments is controlled individually by a $U^{(n)}_\mathrm{seg}$ supplying $\pm40\,\mathrm{V}$, see  \hyperref[fig:1]{Fig. \ref{fig:1}(c)}. We operate the RF blades at $\Omega_\mathrm{RF}=2\pi\cdot17.85\,\mathrm{MHz}$ drive frequency with a peak-to-peak $U_\mathrm{pp}\approx$150$\,\mathrm{V}$ resulting in a radial trapping frequencies of $\omega_{(x,y)}\approx2 \pi\cdot 500\,\mathrm{kHz}$. Micro-motion compensation electrodes are placed behind the RF and DC electrodes (not shown in \hyperref[fig:1]{Fig. \ref{fig:1}}). To position the ion in the trap center, thus compensating residual micro-motion, a DC field is applied in both $x$- and $y$-direction. In axial ($z$-) direction a pair of hollow (inner hole diameter  $400\,\upmu\mathrm{m}$) endcaps are placed at 1.45 mm distance from the trap center. Axial trapping is provided by application of $U^{(6)}_\mathrm{seg}=-0.6\,\mathrm{V}$ on the middle segment and $U^{(n\neq6)}_{\mathrm{seg}}=0\,\mathrm{V}$, resulting in an axial trapping frequency $\omega_z\simeq2\pi\cdot150\,\mathrm{kHz}$. Along the $z$-direction, ions are transported and extracted, as depicted in \hyperref[fig:1]{Fig. \ref{fig:1}(b)}. 

Photoionization loading is achieved using a beam of neutral calcium excited with laser light at $423\,\mathrm{nm}$ and $375\,\mathrm{nm}$ at the trap center. The Ca$^+$ dipole transition $4\text{S}_{1/2}-4\text{P}_{1/2}$ at $397\,\mathrm{nm}$ serves as a Doppler cooling transition. Fluorescence at this wavelength is collected by a lens system with a magnification of $11.6(2)$ and imaged onto an electron multiplier gain charged coupled device (EMCDD) camera, not shown in Fig.~\ref{fig:1}. A laser beam at $866\,\mathrm{nm}$ is required to empty the metastable $3\text{D}_{3/2}$ level. Both beams are aimed at the ion under a direction such that all vibrational modes are laser cooled. 

The trapping module is connected to an extraction and focusing module (ii), denoted with E and F, which is instrumental in controlling the free-space ion propagation. For this we may apply or switch DC voltages on E and F up to $\pm20\,\mathrm{kV}$, independently and within less than $100\,\mathrm{ns}$ time scale. The direction of the ion trajectory requires fine adjustments, despite the high mechanical alignment precision of the modules with respect to each other of $\leq 50\,\upmu\mathrm{m}$. To adjust both the angle and the position, in both radial ($x$-, and $y$-) directions, we developed an (iii) electrostatic steering assembly which is composed of eight half-spherical electrodes (radius of curvature $7\,\mathrm{mm}$, radial distance $12.2\,\mathrm{mm}$). Downstream, a secondary electron multiplier (SEM) can optionally  be used to detect single ions \cite{jacob2016transmission}. 

Triggering of RF and DC electric waveforms is performed using computer controlled switches \cite{kaushal2020shuttling, stopp2021single}. The EMCCD signal is recorded by the same computer. In the experiments presented here, we extract ions from the trap (i), and use the elements (ii) and (iii) such that the ions trajectory is reversed and the ions are recaptured in the trap (i). We name this operation \textit{ion fountain} mode.      

\section{Operation of the ion fountain}
The experimental sequence starts with the ion loading and verification, using a EMCCD picture ensuring exactly one ion is trapped. Then, the extraction DC voltage $U_\mathrm{E}$ and the RF drive voltage $U_\mathrm{RF}$,  respectively, are switched on and ramped down. The endcap voltage is adjusted for extraction, while constant control voltages in the focusing and in the steering module direct the ion trajectory. Both, the timings and the amplitude of DC and RF voltages require a fine optimization. Once the ion has reentered the trapping module via the hole in the endcap, we switch off $U_\mathrm{E}$ and ramp up the RF drive back to ideal trapping conditions. 

A large set of parameters has to be optimized, including the trapping potential in $(x,y,z)$-direction, characterized by $\omega_{(x,y,z)}$ and controlled by the RF and axial trap voltages on the segments $U_\mathrm{seg}^{(n)}$, the extraction voltage(s) $U_\mathrm{E}^{(1,2)}$ on both the endcaps and their timings and pulse duration $\varDelta t_\mathrm{puls}^{(1,2)}$, the focusing voltage $U_\mathrm{F}$ applied to module (ii) and its timing $t_\mathrm{RF}$, the ramping of the RF amplitude $U_\mathrm{RF}(t)$ in phase-stable manner with respect to the endcap pulse timing set by $\varDelta t_\mathrm{puls}$, and steering module control voltages $U_{(x,y)}^{(1,2)}$ for both radial directions $x$ and $y$, respectively.

This large number of parameters, and their large possible ranges, in combination with the lack of gradual feedback, because of the  {\it single ion be-or-not-to-be signal} (either the ion is re-trapped or lost), and the low information acquisition rate make numerical simulations essential before any successful experimental implementation.

\begin{figure}[htbp]
\includegraphics[width=0.49\textwidth]{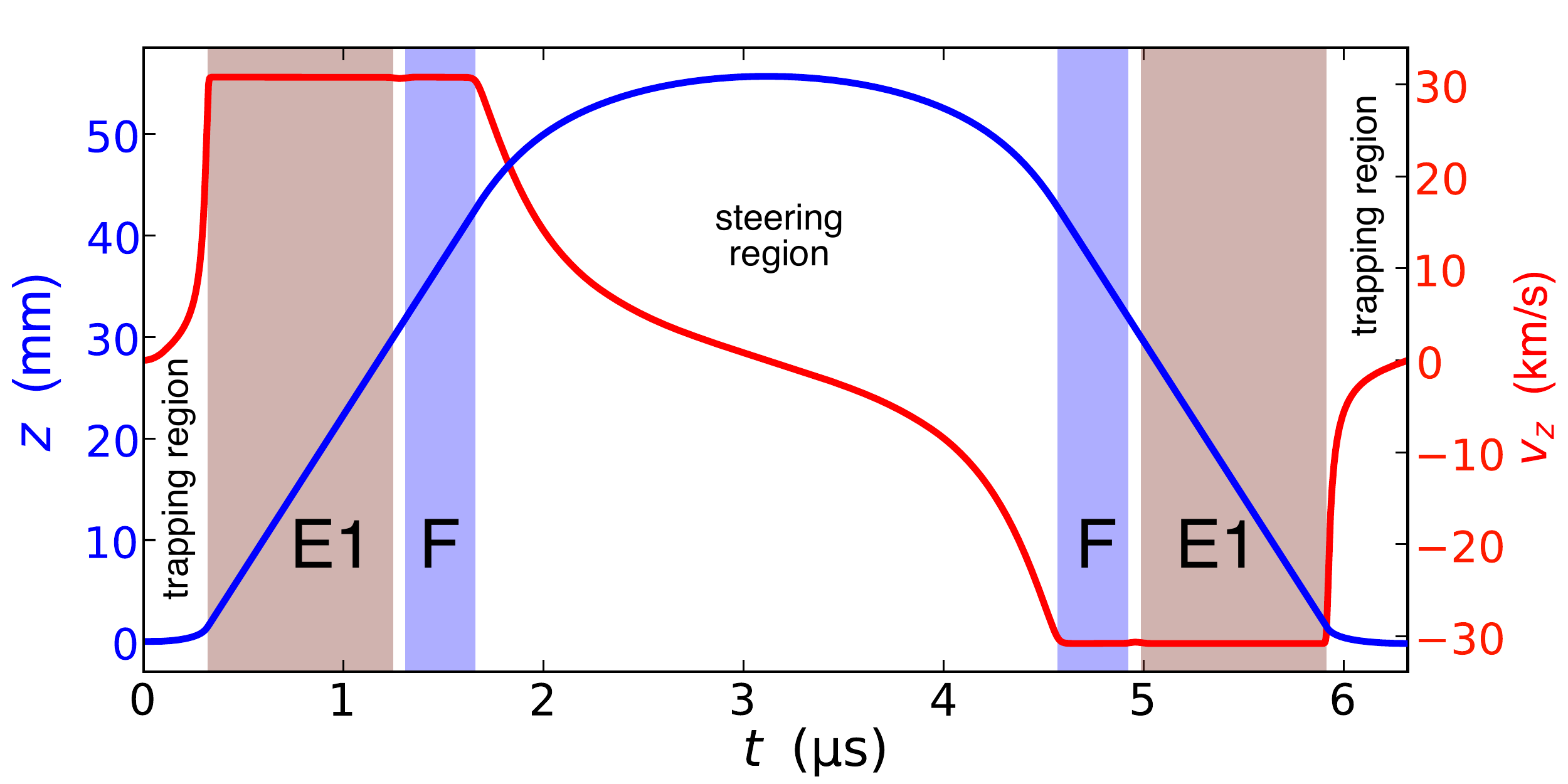}
\caption{Numerical determined trajectory (blue) and velocity (red) of a single $^{40}$Ca$^{+}$ ion along the axial direction. The initial parameters are $v_0=z_0=0$. The entire time-of-flight is $6.3\,\upmu\text{s}$.
}
\label{fig:2}
\end{figure}

\subsection{Numerical simulations for yielding an initial set of operation parameters}

We import CAD geometries \footnote{SOLIDWORKS 2015, https://www.solidworks.com} of the modules (i) to (iii) in the finite element method simulation tool Comsol \footnote{COMSOL Multiphysics, https://www.comsol.com} ($50\,\upmu\mathrm{m}$ grid size) and perform a 1D-simulation of the particle trajectory in the axial potential using a Verlet algorithm. The particle is initially positioned at the trap center and with an initial velocity zero. The ion trajectory is run through in $2\,\mathrm{ns}$ time steps. An event is considered as a successful recapture, if the end of the trajectory is finally found in a region $\leq 100\,\upmu \mathrm{m}$ around the trap center and with velocity $\leq 50\,\mathrm{m/s}$.

We chose to limit our simulations to the most basic procedure with a single negative DC pulse $U_\mathrm{E}^{(1)}$ switched, using identical timing and voltage $U_\mathrm{F}$ also on the focusing module, while $U_\mathrm{E}^{(2)}$ is kept at $0\,\mathrm{V}$. Obviously, the multi-parameter space is excessively large and we only started it's exploration. For the simulation in 1D here, the deflection electrodes are used simply to generate a reflective potential by setting all to one constant voltage $U_{(x,y)}^{(1,2)}=U_\mathrm{R}$. For the initial and the final trapping potential we chose them to be identical with trap frequencies of $\omega^\mathrm{init}_{z}$=$\omega^\mathrm{final}_{z}=2\pi\cdot147\,\mathrm{kHz}$, using $U_\mathrm{seg}^{(6)}=-0.6\,\mathrm{V}$ and $U_\mathrm{seg}^{(n\ne 6)}=0\,\mathrm{V}$, fully in a static fashion. The adjustable parameter of the simulation is the extraction pulse length $\varDelta t_\text{puls}$. It is expected that lower extraction voltages lead to lower kinetic energy of the ion and longer travel time. We assumed that very low extraction voltages would make the process increasingly susceptible, e.g. against slight parasitic charging of the surfaces. For a fixed value of extraction potential $U_\mathrm{E}^{(1)}=U_\mathrm{F}=-200\,\mathrm{V}$, we run the simulation and find a suitable combination of values for $U_\mathrm{R}$, and $\varDelta t_\mathrm{puls}$.

For this extraction voltage and pulse width of $~\varDelta t_\mathrm{puls}^{(1)} =6.3\,\upmu\mathrm{s}$ with $U_\mathrm{R}=7.5\,\mathrm{V}$, we plot the potential and the particle trajectory in  \hyperref[fig:2]{Fig.~\ref{fig:2}}. The particle travel range of $55\,\mathrm{mm}$ reaches the first pairs of semi-spherical deflector electrodes in the steering module, see  \hyperref[fig:1]{Fig.~\ref{fig:1}(b)}. Note, the transport time is only $3\,\upmu\mathrm{s}$ for this travel, which is even below speed for ultra-fast $280\,\upmu\mathrm{m}$ distance segment-to-segment transports \cite{bowler2012coherent, walther2012controlling} by a factor of more than 200.

\subsection{Optimization ion steering}

\begin{figure}[t!]
\includegraphics[width=0.49\textwidth]{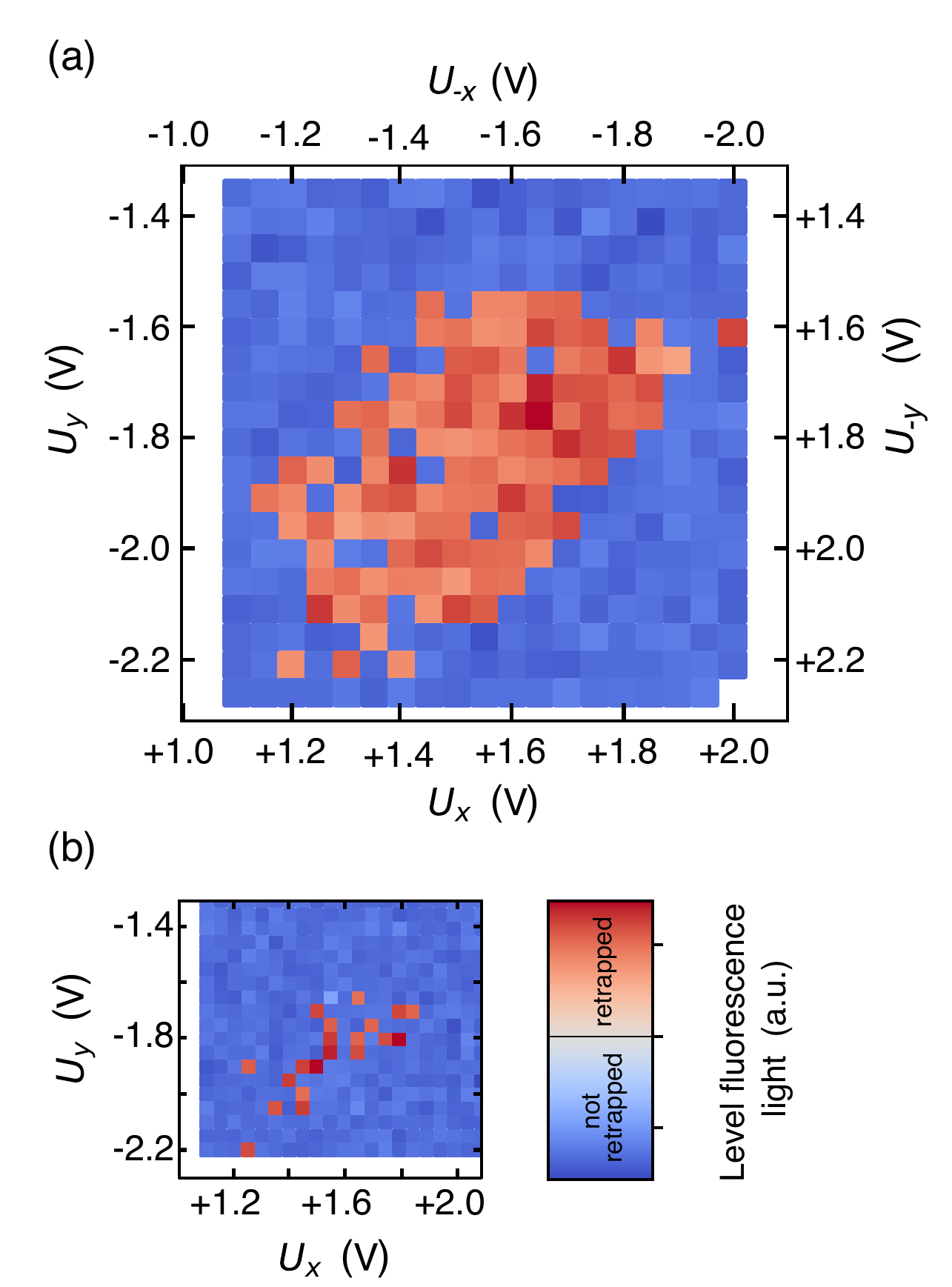}
\caption{Experimental optimization the ion trajectories back reflected into the trap center by scanning $U_{(\pm x,\pm y)}$ relative to $U_\mathrm{R}$. The trapping is verified by checking for laser-induced fluorescence light, for each pixel the ion is retrapped (red) or not (blue). Each pixel corresponds to one single ion events. a) With correct focusing by the reflector for $U_\mathrm{R} =7.5\,\mathrm{V}$, we observe a clear contrast in the retrapping signal. (b) Suboptimal reflection with $U_\mathrm{R}=7.3\,\mathrm{V}$. The scanning range at which the ion can be retrapped is reduced.}
\label{fig:4}
\end{figure}

Using a pre-optimized set of parameters from the numerical simulations with $U_\mathrm{E}^{(1)}=U_\mathrm{F}=-195\,\mathrm{V}$, and  $U_\mathrm{E}^{(2)}=0\,\mathrm{V}$
we start the experimental optimization by scanning deflection voltages $U_{-x}$, $U_{x}$, $U_{-y}$, $U_{y}$ on module (iii) asymmetrically for opposite pairs with $U_\mathrm{R}=7.5\,\mathrm{V}$ to steer the return ion trajectory in $x$- and $y$-direction. This is necessary due to the micro-motion compensation field and imperfections of the trap geometry. For each extraction event, retrapping is verified by waiting one second and checking for single-ion laser-induced fluorescence light with a $30\,\mathrm{ms}$ exposure time on the EMCCD, see \hyperref[fig:4]{Fig. \ref{fig:4}(a)}. We can clearly discriminate whether an ion has been retrapped and is Doppler cooled, or if it is lost. We investigate two-dimensional steering scans for different reflection voltages $U_\mathrm{R}$ between $7.3\,\mathrm{V}$ up to $8.1\,\mathrm{V}$. Note, that for the setting of $U_\mathrm{R}=7.5\,\mathrm{V}$ the scan range of successful deflection voltages is maximized, both in $x$- and in $y$-direction in contrast to \hyperref[fig:4]{Fig. \ref{fig:4}(b)}. We conclude that this setting realizes a cat's eye reflector in time and space for ions. 

\begin{figure}[h!]
\includegraphics[width=0.5\textwidth]{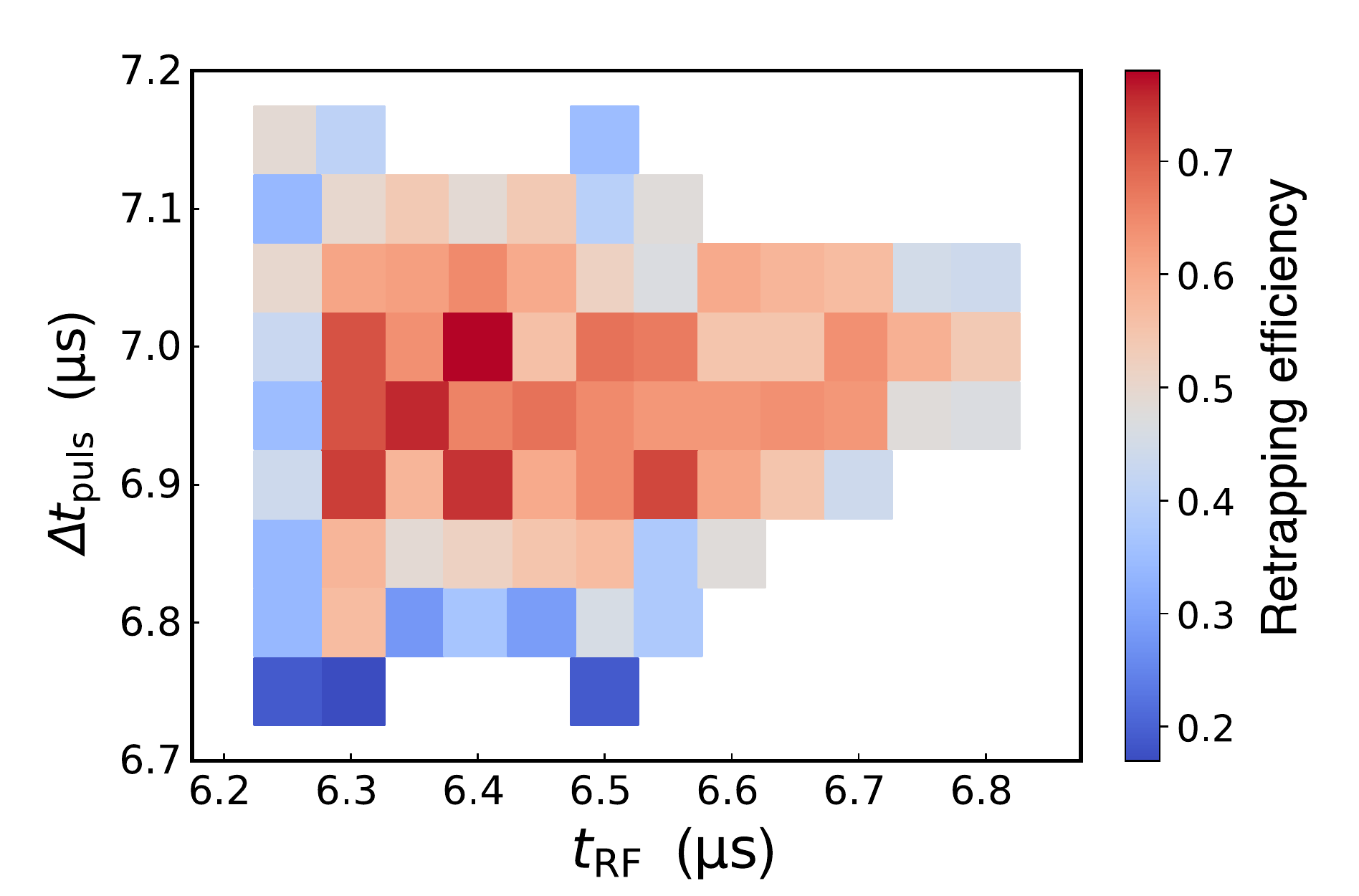}
\caption{Recapture measurement for $t_\mathrm{RF}$ and $\varDelta t_\mathrm{puls}$. For each data point, we average over 100 single ion extraction and recaptures. Optimized timings are at $\varDelta t_\mathrm{puls}=6.95\,\upmu\mathrm{s}$ and $t_\mathrm{RF}=6.35\,\upmu\mathrm{s}$.}
\label{fig:3}
\end{figure}

\subsection{Optimization ion extraction time and determination of the retrapping probability}
Closely related is the optimization of the extraction pulse time $\varDelta t_\mathrm{puls}$ and switch-on time $t_\mathrm{RF}$ of the RF drive, which have to be synchronized to the instant when the ion arrives in the trap. For this a single calcium ion was extracted 100 times, with the voltage pulse $U_\mathrm{puls}$ and the constant RF phase $\varphi_\mathrm{RF}=\Omega_\mathrm{RF}\cdot t_\mathrm{off}$ varied in any combination in steps of $50\,\mathrm{ns}$. The retrapping probability was characterized using the laser-induced fluorescence signal, obtained after a 1~s waiting time. We find the optimal values near $\varDelta t_\mathrm{puls}=6.95\,\upmu\mathrm{s}$ and $t_\mathrm{RF}=6.35\,\upmu\mathrm{s}$, see \hyperref[fig:3]{Fig. \ref{fig:3}}. Note, that the value of $\varDelta t_\mathrm{puls}$ is significantly different compared with the numerical simulation which yields $6.3\,\upmu\mathrm{s}$. This deviation may be caused by errors in the simulation, dominated by effects of the dynamical Paul potential with an oscillating RF field which has not been taken into account. In a linear Paul trap, the RF force at the trap center has no projection in $z$-axis, but we have shown that the time-varying electrical force has a strong $z$-component near the endcap. This leads to an acceleration of the ion, and consequently a change in the overall time-of-flight to return. Even more importantly, a fine-tuning of the RF phase allows for squashing or stretching out the initial velocity distribution, see Ref.~\cite{stopp2021single}.  As a result, the RF phase setting influences the acceptance range for successful retrapping. From the data we estimate a range of about $200\,\mathrm{ns}$. The experimental outcome shows that even though an accurate timing is key, switching at this time scales is unproblematic with this setup.

To obtain an accurate value of the recapture success probability, we used the previously optimized values. A single ion was loaded, which was confirmed using fluorescence light during the Doppler cooling, then extracted and retrapped. The procedure was repeated 752 times and retrapping was successful in 715 cases, resulting in a probability of 0.951. Note, that we typically used the very same single ion over and over again, to avoid time consuming photoionization loading. In one of the runs, the same ion was transported 65 times out of the trapping region and back. The overall data acquisition time was about $30\,\mathrm{min}$, dominated by loading times and the 1~s waiting time. Note, that we have chosen a conservative criteria for retrapping and some of the 37 non-successful events may be due to a recapture in a highly excited motion which could lead to ion loss within the $1\,\mathrm{s}$ waiting time. Also, we observe an ion loss rate due to background gas ($p~\simeq~2\times10^{-9}\,\mathrm{mbar}$) collisions in the order of 1 per minute which would explain about 10 out of the 37 non-successful events.    

\section{Conclusion and outlook}
We experimentally demonstrated a single ion fountain with near-unity recapture efficiency. In the future, we will focus on the remaining challenges towards ion-transport based interconnection of distant quantum processors \cite{lekitsch2017blueprint}: we will use the segmented ion trap with control dynamically of all voltages $U_\mathrm{seg}^{(n)}$ to separate one ion from an ion qubit register, shuttle this single ion to the outermost segment \cite{ruster2017entanglement,kaushal2020shuttling} and extract it from there, while the other ions remain in the linear crystal. For this, we will reduce the susceptibility of the parameters of the extraction from the RF drive by largely reducing the extraction voltage $U_\mathrm{E}^{(1)}$ and controlling F in the focusing module (ii) independently of $U_\mathrm{E}$. The potential from the focusing electrode F is fully shielded from the trapping region. Retrapping parameters will be refined using the well-established Doppler recooling method \cite{wesenberg2007fluorescence, huber2008transport} that allows for an estimate of the motional excitation of the incoming ion once it is trapped. The higher the motional excitation, the longer it takes to achieve full fluorescence level under Doppler cooling light. Ultimately, one might implement sympathetic cooling the incoming ion with another ion species such that the transported qubit ion is ready for high fidelity gate operations. We aim for running teleportation of qubits \cite{riebe2004deterministic,wan2019quantum} between different spatially separated traps and a teleporation of two-qubit gate operations \cite{nielsen2002quantum,gottesman1999demonstrating,riebe2008deterministic}.

Additionally, our method paves the way for quantum sensing applications, where the unique advantages of trapped ions as a sensor \cite{ ludlow2015optical,ruster2017entanglement,brewer2019al+} could be extended to sensing with ions in free space. We envision that the transport of exotic ion species \cite{schmoger2015coulomb}, or isotopes of thorium \cite{groot2019trapping}, or anti-hydrogen ions \cite{indelicato2014gbar}, between different traps plays an important role for experiments that confirm our understanding of the fundamental symmetries of the standard model. Additionally, the single ion extraction may be combined with spin-dependent (light \cite{ monroe1996schrodinger,sorensen1999quantum} or magnetic gradient \cite{khromova2012designer, welzel2018spin}) forces to realize beam splitters \cite{henkel2019stern} or interferometer devices in free space \cite{zych2011quantum, asenbaum2017phase}.      

We thank B. Lekitsch for careful reading the manuscript. We acknowledge financial support by the Bundesministerium für Bildung und Forschung via Q.Link.X., the Volkswagen Stiftung and the Deutsche Forschungsgemeinschaft through the DIP program (Grant Schm 1049/7-1).

\bibliography{bibliography.bib}

\end{document}